\documentclass[11pt,twoside]{article}

\usepackage{asp2006}
\usepackage{epsf}
\usepackage{psfig}
\usepackage{lscape}
\usepackage{graphicx}

\def\DMAG{${\Delta M_{\rm min}}$}
\def\CZ{${\Delta({\rm c} z_{\rm max}})$}
\def\DMAX{${D_{\rm max}}$}

\begin{document}
\title{On The Nature of Fossil Galaxy Groups}   
\author{La Barbera, F. \altaffilmark{1},  
	de Carvalho, R.R. \altaffilmark{2},
	de la Rosa, I.G. \altaffilmark{3},
	Sorrentino, G. \altaffilmark{1},
	Gal, R.R. \altaffilmark{4},
	Kohl-Moreira, J.L. \altaffilmark{5}
}   
\affil{$^{(1)}$INAF-OAC, Via Moiariello 16, 
80131 Napoli, Italy}
\affil{$^{(2)}$INPE/DAS Av. dos Astr. 1758, S\~ao Jos\'e dos Campos, SP Brazil}
\affil{$^{(3)}$IAC, Tenerife, Spain}
\affil{$^{(4)}$UH-IfA, 2680 Woodlawn Dr., Honolulu, HI 96822}
\affil{$^{(5)}$ON, Rua General Jos\'e Cristino 77, Rio de Janeiro Brazil}    

\begin{abstract} 
We present a new sample of 25  fossil groups (FGs) at $z < 0.1$, along
with  a control  sample  of seventeen  bright  ellipticals located  in
non-fossil systems.   Both the global  properties of FGs  (e.g.  X-ray
luminosity)  as well  as the  photometric properties  (i.e.  isophotal
shape  parameter,  $a_4$)  and  spectroscopic  parameters  (e.g.   the
$\alpha$-enhancement) of their first-ranked ellipticals are consistent
with those of the control sample.  This result favors a scenario where
FGs are not a distinct class  of systems, but rather a common phase in
the  life  of  galaxy  groups.   We  also  find  no  evidence  for  an
evolutionary sequence  explaining the formation of  galaxies in fossil
systems through the merging of galaxies in compact groups.
\end{abstract}


\section{Introduction}
A  fossil group consists  of an  isolated, luminous  early-type galaxy
embedded in  an extended  X-ray halo.  The  origin of such  systems is
still a  matter of debate, with  three main scenarios  proposed in the
literature. First, FGs  may be the end-product of  merging galaxies in
normal loose  groups at high redshift  (Ponman et al.   1994; Jones et
al.  2000; Khosroshahi, Jones \& Ponman 2004). The main motivation for
this  idea is that  the merger  time scales  for $\rm  L >  L^*$ group
galaxies are much shorter than the cooling time scales for the hot gas
component in  which they  are embedded (e.g.   Barnes 1989,  Ponman \&
Bertram 1993). In  this simplistic view, FGs can be  used to trace the
mechanisms    driving   the   coalescence    of   galaxies    in   the
not-so-high-redshift  Universe. Alternatively,  FGs  may originate  in
regions that  inhibit the formation  of $L^*$ galaxies, leading  to an
atypical galaxy  luminosity function (Mulchaey \&  Zabludoff 1999). An
intermediate   picture  is  also   plausible,  whereby   the  atypical
luminosity  function of FGs  is a  transient yet  common phase  in the
evolution of groups, ending with the infall of fresh galaxies from the
surroundings~\citep{vonBenda:08}.   In  this  paper,  we  compare  the
properties  of bright  ellipticals in  fossil and  non-fossil systems,
homogeneously selected  from the same dataset, together  with those of
galaxies in Hickson Compact Groups  (HCGs). The FG sample is described
in  Sec.~1,  while  Sec.~2  compares  the properties  of  the  fossil,
non-fossil, and HCG samples. Conclusions are drawn in Sec.~3.

\section{The Fossil Group sample from SDSS+RASS}

We  select FGs using  spectroscopy and  photometry from  SDSS-DR4, and
RASS X-ray  imaging.  We define a volume-limited  sample consisting of
all  galaxies in SDSS  with r-band  Petrosian magnitude  $M_r<-20$ and
spectroscopic    redshift    in     the    range    of    $0.05$    to
$0.095$~\citep{SOR:06}.   A  galaxy  is   defined  as  an  FG  optical
candidate if  it has early-type  morphology and no  companion galaxies
within a magnitude gap of~\DMAG$=1.75$~mag inside a cylinder (parallel
to   the   line-of-sight)   with   a   radius   of   ~\DMAX$=0.35   \,
h_{75}^{-1}$~Mpc   and   a   semi-height   of   ~\CZ$=300$~km/s   (see
~\citealt{LdC:09a}, hereafter LdC09,  for details). We further exclude
galaxies  hosting an  AGN,  classified using  the diagnostic  diagrams
of~\citet{Kew06},  as  well  as  FG  candidates closer  than  $1.5  \,
h_{75}^{-1}$~Mpc to a rich ($R>0$) Abell cluster.
\begin{figure}[!t]
\begin{center}
\includegraphics[width=130mm]{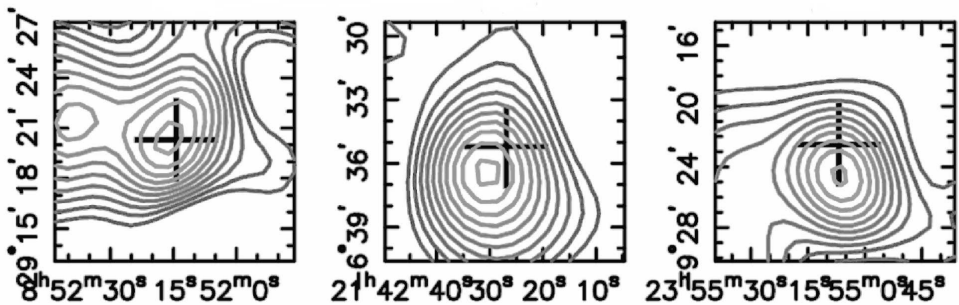}
\caption{Examples  of  X-Ray  contours  around the  optical  positions
  (crosses) of FG first-ranked galaxies (from LdC09).~\label{f1}}
\end{center}
\end{figure}
For  each optical  candidate, we  measure the  X-ray  luminosity $L_X$
($0.5-2.0$~keV)  from the  RASS.  A  source is  considered to  have an
X-ray counterpart  if there  is an X-ray  detection whose  position is
matched  to that  of  the  optical source  within  $1$~FWHM and  whose
luminosity is  3~$\sigma$ above background.  We classify  the FG X-ray
sources as extended if the extension parameter is greater than zero at
the  2$\sigma$   level.   The  extension  parameter   is  measured  by
subtracting in  quadrature the FWHM of  the RASS PSF from  the FWHM of
the FGs.   Our final sample  consists of 25 FGs.   Fig.~\ref{f1} shows
three FG examples of X-ray  contours from the RASS, with the positions
of the optical sources overlaid.

\section{Bright ellipticals in fossil vs. non-fossil systems}
\begin{figure}[!t]
\begin{center}
\includegraphics[width=100mm]{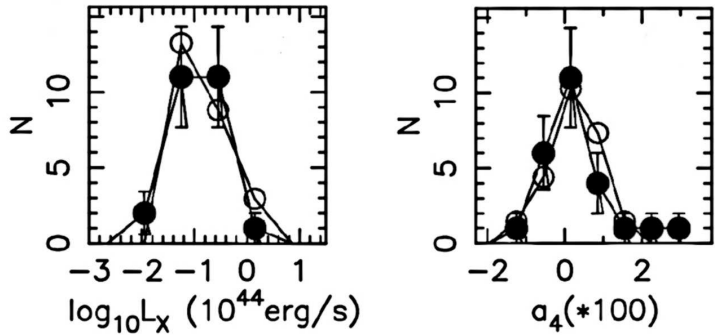}
\caption{Distribution of  $L_X$ (left  panel) and $a_4$  (right panel)
  for  FGs  (filled circles)  and  control  sample ellipticals  (empty
  symbols; normalized  to the  total number of  FGs).  Error  bars are
  1~$\sigma$ Poissonian errors (from LdC09).  ~\label{f2}}
\end{center}
\end{figure}

\begin{figure}[!t]
\begin{center}
\includegraphics[width=90mm]{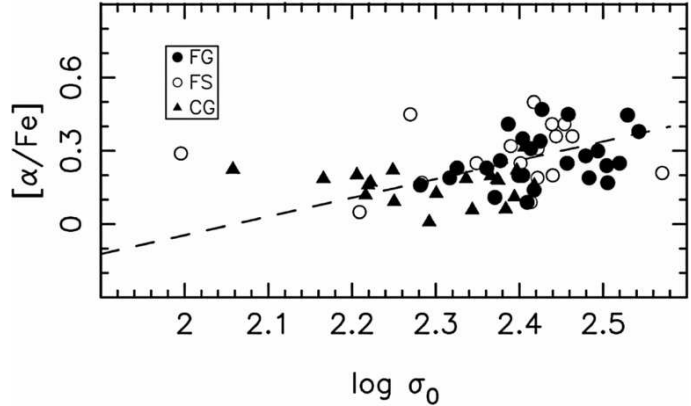}
\caption{The  $\alpha/Fe$ is  plotted  as a  function  of the  central
  velocity  dispersion, $\sigma_0$, for  first-ranked galaxies  in FGs
  (filled circles); ellipticals in the control sample (empty circles);
  and ellipticals in HCGs (filled  triangles).  The dashed line is the
  best-fit line for field galaxies (see LdC09).~\label{f3}}
\end{center}
\end{figure}

We select a  control sample (hereafter CS) of  bright ellipticals from
SDSS+RASS following  the same  procedure as for  the FGs,  but without
applying  the   cylinder  test   for  bright  companions   around  the
first-ranked  galaxy.  This  yields a  sample of  $17$  CS ellipticals
(against  $N=25$  FGs). Fig.~\ref{f2}  compares  the distributions  of
$L_X$ and $a_4$ (describing the deviation of the galaxy isophotes from
purely  elliptical shapes)  for  the  FG and  CS  systems.  The  $L_X$
distributions are fully consistent (using the KS test).  Since FGs and
CS  ellipticals have similar  optical luminosities  (by construction),
the consistency  of $L_X$  implies that fossils  do not  have enhanced
X-ray luminosity  compared to non-fossil galaxies, in  contrast to the
results   of~\citet{Khosroshahi:07}   and~\citet{Jones:03},   but   in
agreement  with  ~\citet{Voevodkin:09}.   The distributions  of  $a_4$
values are  also consistent, with  FG and CS galaxies  presenting both
disky  ($a_4>0$) and  boxy  ($a_4<0$) isophotes.   This  result is  in
disagreement with  that of~\citet{Khosroshahi:06}, who  concluded that
FG systems have  preferentially boxy isophotes (for a  sample of seven
FGs).  On  the other  hand, the  fact that FG  ellipticals can  have a
variety  of isophotal  shapes agrees  with the  predictions  of N-body
simulations from~\citet{Diaz:08}.
We also investigate the stellar population properties of the FG and CS
samples  - age,  metallicity, and  $\alpha$-enhancement, $[\alpha/Fe]$
(see LdC09  for details).  The  twenty elliptical galaxies  in Hickson
compact groups (HCGs),  studied by de la Rosa  et al.~(2007), are also
included  in  the analysis.  Fig.~\ref{f3}  shows  $[\alpha/Fe]$ as  a
function of the central  velocity dispersion, $\sigma_0$, for galaxies
in the three samples.  We  find that the stellar population properties
of  FG galaxies are  very similar  to those  of bright  ellipticals in
non-fossil systems while ellipticals  in HCGs have lower $[\alpha/Fe]$
relative to both FG and CS galaxies.

\section{Conclusions}
We find a striking consistency between fossil and non-fossil galaxies.
This  similarity  indicates that  FGs  are  not  a separate  class  of
systems, but represent  a common phase in the  evolutionary history of
galaxy  groups, which can  undergo fossil  phases at  different cosmic
epochs, with these  phases ending with an infall  of gas rich galaxies
from their surroundings~\citep{vonBenda:08}.  Our results suggest that
generally  there is no  evolutionary link  between fossil  and compact
group (CG) galaxies.  In fact, dry mergers of ellipticals in CGs would
not increase their $\alpha$-enhancement, and cannot therefore generate
the  higher $\alpha/Fe$  observed  in  most of  the  FG galaxies  (see
Fig.~\ref{f3}). On the other hand, wet mergers also seem not to play a
major  role,  since  we  do   not  find  any  excess  of  first-ranked
ellipticals  with either  disky isophotes  or positive  internal color
gradients (see LdC09). However,  if ellipticals in CGs at intermediate
redshifts  have different  properties from  those  in CGs  at $z  \sim
0$~\citep{dC:05,  Mendes:07},   our  results  can  not   rule  out  an
evolutionary sequence between the higher-$z$ CGs and low-$z$ FGs.


\acknowledgements  
We acknowledge the use of  SDSS  data. \\
(see http://www.sdss.org/collaboration/credits.html)


\end{document}